\documentclass[11pt,oneside]{article}
\usepackage[english]{babel}
\usepackage{amsmath,amsfonts,amssymb}%,amsthm
\usepackage[T1]{fontenc}
\usepackage{graphicx}
\usepackage{bm}
\usepackage{tikz}
\usepackage{booktabs}
\usepackage{multirow}
\usepackage[TABBOTCAP]{subfigure}
\usepackage{cases,soul}
\usepackage[letterpaper,margin=0.9in]{geometry}
\usepackage{soul}

\newcommand{\beq}{\begin{equation}}
\newcommand{\eeq}{\end{equation}}

\usepackage{tabularx}
\usepackage{xcolor}
\definecolor{blun}{cmyk}{0.8, 0.5, 0, 0.7}
\definecolor{darkpastelred}{rgb}{0.76, 0.23, 0.13}
\definecolor{bostonuniversityred}{rgb}{0.8, 0.0, 0.0}

\usepackage[bookmarks,colorlinks,pdfauthor={Dr. Cinzia Soresina},linkcolor=blun,citecolor=blun,urlcolor=blun]{hyperref}

\let\oldbibliography\thebibliography
\renewcommand{\thebibliography}[1]{%
  \oldbibliography{#1}%
  \setlength{\itemsep}{-1.2mm}%
}

\begin{document}

\title{Mortality estimate driven by population abundance field data in a stage-structured demographic model. The case of \emph{Lobesia botrana}}

\author{
S. Pasquali\thanks{CNR-IMATI ``Enrico Magenes'', via Alfonso Corti 12, 20133 Milano, Italy - \href{mailto: sara.pasquali@mi.imati.cnr.it}{sara.pasquali@mi.imati.cnr.it}}, 
C. Soresina\thanks{Institut f\"ur Mathematik und Wissenschaftliches Rechnen, Universität Graz, Heinrichstr.~36, 8010 Graz, Austria - \href{mailto: soresina@ma.tum.de}{cinzia.soresina@uni-graz.at} - ORCiD: 0000-0002-0247-3632}, E. Marchesini\thanks{AGREA S.r.l.~Centro Studi, via Garibaldi 5/16, 37057 S. Giovanni Lupatoto (VR), Italy -  \href{mailto: enrico.marchesini@agrea.it}{enrico.marchesini@agrea.it}}}

%\institute{S. Pasquali \at
%	CNR-IMATI ``Enrico Magenes'', via Alfonso Corti 12, 20133 Milano, Italy\\
%           Tel.: +39 0223699525\\
%           \email{sara.pasquali@mi.imati.cnr.it}           
%           \and
%           C. Soresina \at% Technische Universit\"at M\"unchen
%	   Zentrum Mathematik, TUM, Boltzmanstr. 3, 85748 Garching bei M\"unchen, Germany\\
%           Tel.: +49.89.289.18312\\
%	ORCiD: 0000-0002-0247-3632\\
%           \email{soresina@ma.tum.de}
%           \and
%           E. Marchesini \at
%	AGREA S.r.l. Centro Studi, via Garibaldi 5/16, 37057 S. Giovanni Lupatoto (VR), Italy\\
%           Tel.: +39 045548412\\
%           \email{enrico.marchesini@agrea.it}
%}

\maketitle
%%%%%%%%%%%%%%%%%%%%%%%%%%%%%%%%%%%%%%%%%%%%%%%%%%%%%%%%%%%%%
\begin{abstract}\noindent
Simulating the population dynamics of a stage-structured population requires the knowledge of the biodemographic functions characterizing the species, namely development, mortality and fecundity. In general, development and fecundity can satisfactorily be estimated starting from literature data. Unfortunately, this is often not the case of the mortality function, because of the lack of experimental data. To overcome this problem we estimate the mortality rate function from field data on the abundance of the species. The mortality is expressed as a linear combination of cubic splines and the estimation method allows to determine its coefficients taking into account the observations measurement error. Moreover, the variability in the estimate is quantified by means of the confidence bands for both mortality and dynamics. The presented method allows to obtain a more flexible shape for the mortality rate functions compared with previous methods applied to the same pest.
% and the corresponding mortality function with its confidence bands. 
The method is applied to the case of \emph{Lobesia botrana}, the main pest in the European vineyards, with abundance data collected in a location in the North of Italy for five consecutive years. Data collected over three years are used to estimate the mortality and to analyze the variability in the estimate and its effects on the population dynamics. Other two datasets are used to validate the model simulating the dynamics using the estimated mortality. %The dynamics give a satisfactory fit of the phenology of the pest. 

\end{abstract}
%\textbf{AMS Subject Classification (2010):} 35R11, 37M20, 35B32
%%%%%%%%%%%%%%%%%%%%%%%%%%%%%%%%%%%%%%%%%%%%%%%%%%%%%%%%%%%
\section{Introduction}\label{Intro}
Population dynamics models play an important role in pest control. A good knowledge of the temporal dynamics of a pest population can help decision makers in the choice of the best strategy in terms of application of phytosanitary treatments. This is a fundamental task in light of the Directive 2009/128/EC on the sustainable use of pesticides in Europe.

In order to obtain a good description of the population dynamics it is necessary to take into account climatic factors, phenology of the plant and, in general, physical-biological characteristics of the environment in the site of interest.
The population dynamics can be represented using a phenological model that describes the percentage of individuals in the different stages or a physiologically-based demographic model that accounts for the population abundance in time. Effects of mortality and fecundity rate functions in phenological models are discussed in Pasquali et al.~\cite{pasquali2019}.
Physiologically based demographic models allow to know the population abundance over time, taking into account the environmental variables influencing the dynamics of a species. These mechanistic models have been used since a long time (see for instance~\cite{dewit1974, gutierrez1975, mcdonald1989, metzdiekmann1986, wang1977}). 
Most of the pests are dangerous for the crop only in a particular phase of their life, then it is often useful to consider the population organized in stages, giving rise to the so-called stage-structured population models \cite{cushing98,Iannelli94,IannMil10}.
%Most of the pests are dangerous for the crop only in a particular phase of their life, then it is often useful to consider the population organized in stages \textcolor{red}{like in the stage-structured population models \cite{cushing98,Iannelli94,IannMil10}}.
%Often, it is useful to consider the population organized in stages because pests are dangerous for the crop when they are in a particular phase of their life. 
Physiologically-based demographic model have been widely used in the last years to describe pest population dynamics for stage-structured populations~\cite{ainseba2011, blum2018, ewing2016, gilioli2016, gilioli2017, gilioli2014, marini2016, pasquali2020, rossini2020distributed}.
Here we consider a physiologically-based demographic model for a stage-structured population described by a system of partial differential equations. It is a particular case of a more general model presented in~\cite{BuffoniPasquali2007}. The model gives the abundance of the population in each stage in time and physiological age. It is based on biodemographic functions (development, mortality and fecundity) describing the biology of the species. Development, mortality and fecundity rate functions are stage-specific and depend on environmental variables, mainly temperature.

It is important to have a good estimate of the biodemographic functions to obtain a reliable model. In general, these rate functions are estimated starting from literature data on the biology of the species (for instance, duration within a stage for the development function, number of eggs produced by an adult female for the fecundity function). Starting from these data, a simple least square method allows to estimate the parameters of a biodemographic function of a given functional form. Unfortunately, for the mortality function data are often not available in literature. In this case, different methods of estimation can be applied (see for instance~\cite{WoodNisbet1991} for a survey of mortality estimation methods). When laboratory data on the mortality rate are not available, mortality estimate can rely on the knowledge of time series data on population dynamics. Different methods to estimate mortality, starting from population dynamics time series data, have been proposed in the last years. Ellner et al.~\cite{ellner2002} proposed a non-parametric regression model, Picart and Ainseba~\cite{picartainseba2011} solved the problem using a numerical analysis based on a Quasi-Newton method, in~\cite{gilioli2016} a method based on least squares was presented, while in~\cite{lanzarone2017} a Bayesian estimation method was described. In the two last approaches a functional form for the mortality is required and the estimate concerns only parameters appearing in this function. This can be restrictive in some cases, where having more flexibility for the mortality shape is preferable. Then we decided to follow the approach proposed by Wood~\cite{wood2001} which does not require a functional form, but rather it expresses the mortality as linear combination of elements  of a suitable basis. %In particular, we choose a basis of cubic splines.
The parameter vector containing the coefficient of the linear combination is estimated by minimizing a weighted least squares term that measures the ``distance'' between the simulated and the collected population abundance.
%The coefficients of the linear combination form a vector parameter \textcolor{red}{that in the philosophy of~\cite{wood2001} should be}  
Different weights can be considered for the various stages of the population and/or for the various generations.
Here, differently from~\cite{wood2001}, we substitute the functional to be minimized to take into account the measurement error typical of field observations. The new objective is to minimize the sum of the weighted ``distance'' between the simulated abundance and the observations range of variability.
Furthermore, also the variability in the estimation has been taken into account by considering the confidence bands for both the mortality rate functions and the population dynamics. Since the estimator of the parameter vector is a random variable, confidence bands are obtained by drawing a certain number of values for the parameter vector from its distribution.\\
\indent
The mortality estimation method is applied to the case study of the grape berry moth {\em Lobesia botrana} which is considered the most dangerous pest in European vineyard. 
%\st{Once estimated the mortality rate functions for all the stages, it is possible to simulate the dynamics of the pest.}
The problem of the description of the grape berry moth dynamics has been widely studied in the past \cite{ainseba2011,baumgartnerbaronio1988,gutierrez2012,gilioli2016}. 
To check the goodness of the estimation method, we apply it to the case of data generated from the population dynamics system with known mortality rate functions. Then the method is applied to a real case. Here we consider a dataset on population abundances of the grape berry moth collected in Colognola ai Colli (Verona, Italy) in the period 2008--2012 for the cultivar Garganega.
The dataset is split into two groups: population abundances of years 2008, 2009 and 2011 are used to estimate the mortality, while data of 2010, and 2012 are used to validate the model.
The proposed method allows to know the behaviour of the mortality rates as function of the temperature. Since the grape berry moth is dangerous only in its larval stage, it is useful to assign a higher weight to this stage when considering the least square term in the estimation procedure. This can be useful to the end of grape berry moth control in vineyards~\cite{picart2011,picart2014,picart2015}. 
% Then, we consider two cases for the mortality estimate: equal weights for all the stages, higher weight for larvae. In the latter case a better fit of the larval dynamics is obtained.
% The variability of the estimate is analyzed for the case of equal weights for all the stages. 
The variability in the mortality rate functions reflects into a variability for the population dynamics. The variability is quantified by means of the confidence bands for both the mortality and the dynamics. 
%\textcolor{red}{The largest uncertainty corresponds to the period of the abundance peaks.}
The estimated mortalities can be used to forecast the population abundance in future periods, or to simulate the grape berry moth dynamics in other nearby locations. The estimation method here described is sufficiently general to be applied to other stage-structured populations.
\medskip

The paper is organized as follows. In Section 2 the mathematical model describing the dynamics of the population is presented, while in Section 3 the biodemographic functions of the grape berry moth are specified. Section 4 describes the mortality estimation method and the method to determine the confidence bands. In Section 5, the estimation method is applied to the grape berry moth for both the case of simulated data and of field data. Finally, Section 6 is devoted to discussion and concluding remarks.

\section{The mathematical model}

The demographic model is based on a system of partial differential equations that allows to obtain the temporal dynamics of the stage-structured population and their distribution over physiological age within each stage. Let
\begin{align*}
\phi^i(t,x) dx =& \textnormal{ number of individuals in stage $i$ at time $t$}\\
                & \textnormal{ with physiological age in $(x,x+dx)$,}
\end{align*}
\noindent
for $i=1,2,...,s$, where $s$ is the number of stages. Stages from $1$ to $s-1$ are immature stages, and stage $s$ represents the reproductive stage (adult individuals). Note that $t$ denotes the chronological time while $x$  represents the physiological age indicating the percentage of development within the stage over time~\cite{BuffoniPasquali2007, BuffoniPasquali2010, BuffoniPasquali2013, diCola1999}.

Instead of a deterministic setting  in which the population dynamics is described through von Foerster equations~\cite{BuffoniPasquali2007}, we prefer to consider a stochastic approach which allows to take into account the variability of the development rate among the individuals~\cite{BuffoniPasquali2010,BuffoniPasquali2013}. The dynamics is described in terms of the forward Kolmogorov equations~\cite{gardiner1986,carpi1988}

\beq
\frac {\partial \phi^i} {\partial t} + \frac {\partial} {\partial x}
\left[ v^i(t) \phi^i - \sigma^i \frac  {\partial \phi^i} {\partial x} \right]
+ m^i(t) \phi^i = 0, \quad t > t_0, \; x \in (0,1),
\label{kolm1}
\eeq

\beq
\left[ v^i(t) \phi^i(t,x) - \sigma^i \frac {\partial \phi^i} {\partial x}
\right]_{x=0} = F^i(t),
\label{kolm2}
\eeq

\beq
\left[ - \sigma^i \frac {\partial \phi^i} {\partial x} \right]_{x=1} = 0,
\label{kolm3}
\eeq

\beq
\phi^i(t_0,x) = {\hat \phi}^i(x),
\label{kolm4}
\eeq

\noindent
where~$i=1,2,...s$,~$v^i(t)$ and~$m^i(t)$ are the specific development and mortality rates, respectively, assumed to be independent of the age $x$; ${\hat \phi}^i(x)$ are the initial distributions, while $\sigma^i$ are constant diffusion coefficients. The boundary condition at $x=0$ assigns the input flux into stage $i$, while the boundary condition at $x=1$ means that the output flux from stage $i$ is due only to the advective component $v^i(t) \phi^i(t,1)$ \cite{BuffoniPasquali2007}. Moreover, the fluxes $F^i(t)$ in the boundary conditions~\eqref{kolm2} are evaluated as follows. The term $F^1(t)$ is the egg production flux and is given by

\beq
F^1(t) = v^s(t) \int_0^1 \; \beta(t,x) \; \phi^s(t,x) \; dx,
\label{vonfos4}
\eeq

\noindent
where $v^s(t) \beta(t,x)$ is the specific fertility rate. In particular, we consider

\beq
v^s(t) \beta(t,x) = b(t) f(x) \; \;  eggs/adults \; with \; age \; in \; (x,x+dx)/time \; unit,
\label{bt}
\eeq

\noindent
where~$b(t)$ takes into account the effect due to both diet and temperature, and $f(x)$ is the fertility profile. 

The other terms~$F^i(t)$, when~$i>1$, are the individual fluxes from stage $i-1$ to stage $i$ and are given by  

\beq
F^i(t) = v^{i-1}(t) \phi^{i-1}(t,1), \quad i>1.
\label{vonfos5}
\eeq

The functions $\phi^i(t,x)$ allow to obtain the number of individuals in stage $i$ at time $t$:
$$N^i(t)=\int_0^1 \phi^i(t,x) dx.$$

System \eqref{kolm1}--\eqref{kolm4} requires an explicit formulation (depending on a certain number of parameters) of basic biodemografic rate functions (development, fecundity and mortality) that models the physiological response of individuals to environmental forcing variables. For the grape berry moth, as for all the poikilotherm organisms, temperature is considered the most important driving variable; then, these biodemographic rate functions are formulated in terms of temperature, which depends on the chronological time. It is also possible to take into account the dependence on other environmental variables ~\cite{Schmidt2003}.

\section{Structure of {\em L. botrana} population}
\label{lobesiabotrana}

\emph{Lobesia botrana} has a stage-structured population, generally considered composed by four stages: eggs, larvae, pupae and adults~($s=4$). 
%Estimations of stage-specific biodemographic functions usually rely on bottom-up laboratory experimental data, while top-down field population data must be used to validate the model. 
Each stage is characterized by its own biodemographic functions (development, mortality, and, for adults, also fecundity). In this section we define development and fecundity rate functions for \emph{L.~botrana} that summarize our knowledge on the biology of the species. We suppose that development and mortality rate functions depend on time only through temperature, while the fecundity rate function depends also on the physiological age, as done in~\cite{gilioli2016}.

Laboratory experimental data are used to estimate developmental and fecundity rates. On the contrary mortality rates are very difficult to measure. For this reason we use population time series data to estimate the mortality applying a modified version of the method proposed by Wood~\cite{wood2001} for formulating and fitting partially specified models.

\subsection{Development rate function}
The development rate function $v(t)$ appearing in~\eqref{kolm1} describes the development response curve. Typically, there is no growth below a lower temperature threshold, while the developmental rate increases and reaches a maximum at an optimal temperature and then it declines rapidly approaching zero at a thermal maximum (temperature at which life processes cannot longer be maintained for prolonged periods of time \cite{logan1976analytic}). %a lethal temperature threshold. 
Several functional expressions have been proposed in literature to describe development \cite{Kontodimas2004}. Here, as in~\cite{gilioli2016}, we consider a Lactin function \cite{lactin1995} to represent the development of all the stages:
\begin{equation}
v(t)= \delta^i \max\left\{0,\textnormal{e}^{\alpha^i T}-\textnormal{e}^{\alpha^i T_m -\frac{T_m -T}{\beta^i}}-\gamma^i \right\}
\label{eq:lactin}
\end{equation}
where $T_m=36 \, ^oC$ is the thermal maximum, $\alpha^i$ is the slope parameter describing the acceleration of the function from the low temperature threshold to the optimal temperature, $\beta^i$ is the width of the high temperature decline zone, $\gamma^i$ 
%is the asymptote to which the function tends at low temperatures
is a parameter that allows the curve to intersect the abscissa giving a minimum temperature of development, and $\delta^i$ is a coefficient of amplification of the curve.

Parameters of the development rate functions are estimated by means of a least square method in \cite{gilioli2016} using the datasets in \cite{baumgartnerbaronio1988,brierepracros1998}, and are reported in Table \ref{TabDev}.
\begin{table}
\centering
\begin{tabular}{ccccc}
\toprule
	    & $\alpha^i$ & $\beta^i$ & $\gamma^i$ & $\delta^i$ \\
\midrule
$i=1$ & 0.01 & 0.8051 & 1.0904 & 1\\
$i=2$ & 0.003 & 0.662 & 1.0281 & 1\\
$i=3,4$ & 0.0076 & 1.7099 & 1.0929 & 1.1\\
\bottomrule
\end{tabular}
\caption{Parameters of the stage-specific development rate function in \eqref{eq:lactin} for the four stages of \emph{L. botrana}: eggs ($i=1$), larvae ($i=2$), pupae ($i=3$) and adults ($i=4$).}
\label{TabDev}
\end{table}

\subsection{Fecundity rate function}
As already supposed in \eqref{bt}, we assume that the eggs production depends both on the physiological age of the adults and on the chronological time through temperature and phenological stage of the host plant as environmental variables. The oviposition profile $f(x)$ in equation \eqref{bt}, as function of the physiological age $x$, is assumed to be of the functional form
$$f(x)=a x^{b-1}\exp{(-cx)},$$
where the parameters $a,\, b,\, c$ are usually estimated from experimental data. 
This class of functions, reproducing the shape of a gamma distribution, is sufficiently general to allow the shift of the mode in all the values of the physiological age interval.

The term $b(t)$ in equation \eqref{bt} takes into account the influence of environmental variables, temperature $T(t)$ and phenological stage of the plant $P(t)$, varying with the chronological time. It is expressed by the product
$$b(t)=b_0\left(P(t)\right)a_0\left(T(t)\right),$$
where $b_0(\cdot)$ is a step function indicating the insect diet changing over time due to the plant maturation process, and 
$$a_0(T)=1-\left( \dfrac{T-T_L-T_0}{T_0}\right)^2$$
captures the effect of temperature. The parameter $T_L$ indicates the minimum temperature of reproduction, while $T_0$ the half-width of the temperature reproduction interval.

Regarding the case study of \emph{L. botrana}, the parameters appearing in the function $f(x)$ of the fecundity rate are obtained fitting the corresponding oviposition profile in~\cite{baumgartnerbaronio1988},  properly converted to a function of the physiological age; their values are
$$a=74270,\quad b=4.06,\quad c=15.33.$$
The values appearing in the function $a_0(T)$ are \cite{gilioli2016,gutierrez2012}
$$T_L=17 \,^oC,\quad T_0=7.5 \, ^oC.$$

\begin{figure}%
\includegraphics[width=0.45\columnwidth]{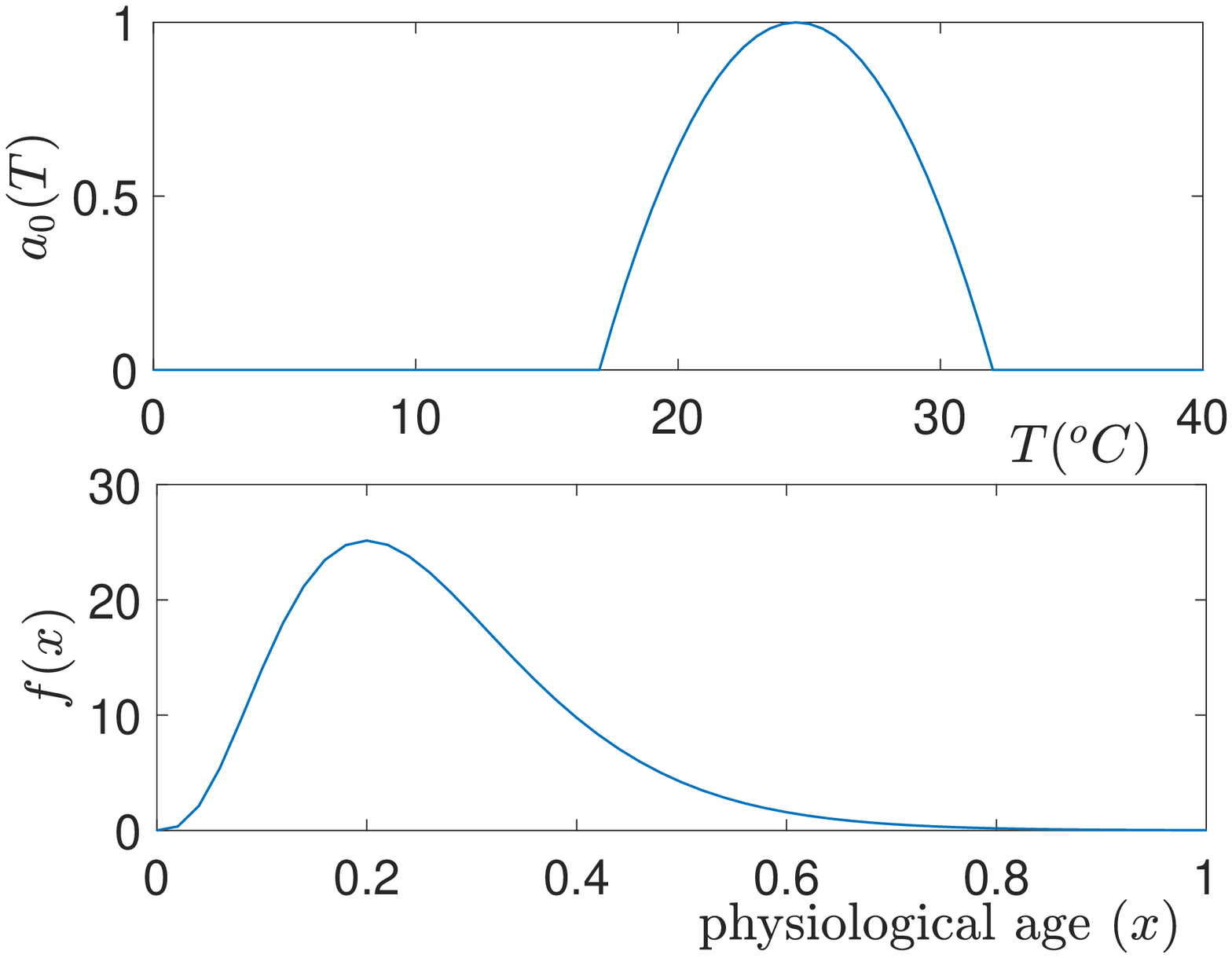}%
\includegraphics[width=0.55\columnwidth]{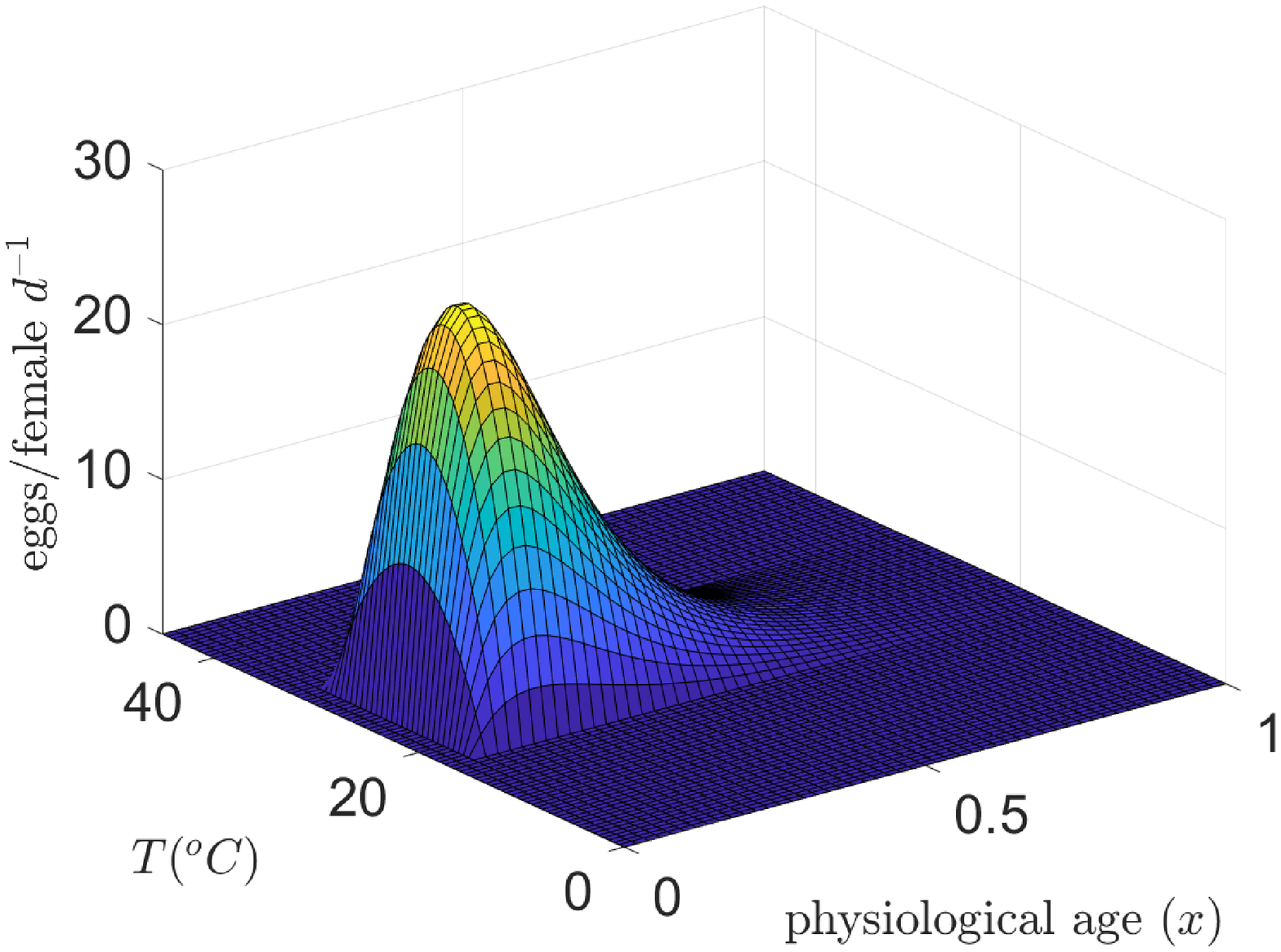}%
\caption{(left panel) Temperature dependent factor and oviposition profile as function of the physiological age. (right panel) Fecundity rate function ($eggs/female \,days^{-1}$) on temperature ($^\circ C$) and physiological age (dimensionless) for the adult stage of \emph{L. botrana} for $b_0(P)=1$.}
\label{FigFec}
\end{figure}

The product $f(x)a_0(T)$ ($eggs/female\, day^{-1}$) over temperature ($^\circ C$) is illustrated in Figure~\ref{FigFec}. Function $b_0(P(t))$, which depends on the phenological age $P$ of the plant expressed in terms of BBCH-scale~\cite{lorenz1994}, is a step function with steps at the BBCH stages indicated in Table~\ref{TabPlant}~\cite{gutierrez2012}. 
\begin{table}
\centering
\begin{tabular}{ccc}
\toprule
Plant stage & $P$ & $b_0(P)$\\
\midrule
Inflorescence & BBCH 53 & 0.31 \\
Green Berries  & BBCH 71 & 0.48\\
Maturing fruits  & BBCH 81 & 1 \\
Berries ripe for harvest& BBCH 89 & 0 \\
\bottomrule
\end{tabular}
\caption{Values of the step function $b_0(P)$, with steps in three plant phenological stages, following the BBCH-scale~\cite{lorenz1994}.}
\label{TabPlant}
\end{table}

\subsection{Mortality rate function}
\label{submort}
%The mortality rate function $m(t)$, also called temperature dependent intrinsec mortality, can be defined starting from the average stage proportional mortality and the stage-specific developmental rate. In \cite{gilioli2016} the authors considered an extrinsic mortality term due to natural control factors, which depends on the stage and on the generation. 

Since mortality rates are very difficult to measure, then the functional form of the mortality rate $m(t)$ cannot be easily determined as the development and the fecundity rate functions. In~\cite{gilioli2016,lanzarone2017} the authors introduced a mortality composed by two terms: an intrinsic temperature-dependent (abiotic) mortality depending on the development rate function, and a constant generation-dependent extrinsic mortality likely related to external natural control factors, to be estimated using time series field data on the population dynamics. 
Ainseba et al.~\cite{ainseba2011} also proposed mortality rate functions of a known form depending on both environmental variables and age.
In this paper we consider the mortality $m(t)$ as unknown and we apply a modified version of the estimation method proposed by Wood~\cite{wood2001} for formulating and fitting partially specified models. Thanks to this approach, the obtained mortality carries out both extrinsic and intrinsic mortality factors.
Moreover, since we want to estimate the mortality rates without assuming a specific functional form for them, we do not specify an analytical expression, but we only consider some biologically meaningful hypothesis:
\begin{itemize}
	\item[-] the mortality rate depends on the chronological time through temperature;\\[-0.7cm]
	\item[-]  the mortality rate is a non-negative continuous function of temperature;\\[-0.7cm]
	\item[-]  the mortality rate is strictly positive at two reference temperatures.\\[-0.7cm]
\end{itemize}
These assumptions will constitute the constraints on the shape of the mortality rates in the sequel.

\section{Estimation of the mortality rate function}
% To estimate the mortality, we apply the method proposed by Wood \cite{wood2001} for formulating and fitting partially specified models. \textcolor{red} {modifying the functional to be minimized to take into account the measurement error}. 
This method allows to deal with a mortality of unknown form guaranteeing more flexibility for its shape than in~\cite{gilioli2016}. 
In order to apply this method we need a data set of population dynamics, meaning observations of the abundance of the different stages over time. Obviously, these observations are subject to a measurement error depending on the stage, because detecting the abundance in some stages is more difficult than for others.
%%%%%%%%
% Firstly, we generate a dataset of abundances for the four stages of the grape berry moth to check the convergence of the method. Then, we consider the same dataset used in \cite{gilioli2016} relative to the dynamics of the grape berry moth 
% in a vineyard of Garganega located in Colognola ai Colli, a hilly region in the North-East of Italy during the period 2008-2012. 
% The experimental field was not treated with insecticides to avoid controls on the growth of the insect population. 
% More precisely, to estimate the mortality rates functions we used the field data collected at Colognola ai Colli in the three years 2008, 2009 and 2011 (model calibration); the data for the other years 2010 and 2012  were used to test the model (validation), keeping all the other parameters of development and fecundity fixed. 
%%%%%%%%%%%%

\subsection{The estimation method}
\label{method}
We consider system \eqref{kolm1}--\eqref{kolm4} in which the development functions and the fecundity function are chosen as in the previous section, and the mortality rates are unspecified. We want to find the functions~$m^i(t)\,(i=1,\dots,s)$ that result in the best fit of the model to field data of populations densities. Once the model functions are fixed, system \eqref{kolm1}--\eqref{kolm4} can be numerically solved and produce a vector of model estimates $\bm{\mu}$, representing the population abundances, corresponding to the observations $\bm{y}$. Usually observations are affected by errors. In particular in our case study it is very difficult to detect insects in some stages of their life, meaning that the observations could underestimate the real abundances. We denote by $\bm{u}$ the vector representing the measurement error. Then, we expect the real abundances included in the interval  $[\bm{y},\bm{y}+\bm{u}]$.
The goodness of the fit can be quantitatively measured as a weighted sum of squared distance of each observation $y_k$ from the interval $[y_k,y_k+u_k]$:
%weighted least squared term
$$\sum_{k=1}^{d} \max\left\{0,(y_k-\mu_k)w_k(y_k+u_k-\mu_k)\right\},$$
where $d$ is the number of data and $w_k$ are the weights associated to each measurement. Then, the best fitting functions $m^i$ are those which minimize this quantity.

The unknown functions $m^i$ can be expressed as linear combination of elements of a suitable basis $\xi_{i,j}(t),\, i=1,\dots,s,\,j=1,\dots,n_i$ (for instance, a polynomial or cubic spline basis)
$$m^i(t)=\sum_{j=1}^{n_i}p_{ij}\xi_{ij}(t).$$  
Thanks to this choice, the mortality is then very general and it is not limited to a fixed form. The problem of finding the best fitting functions $m^i$ is reduced to finding, under some constraints, the best fitting parameters $p_{ij},\; i=1,\dots,s,\;j=1,\dots,n_i$, collected into the vector $\bm{p}=[p_{11},\dots,p_{1n_1},\dots,p_{sn_s}]^T$, which produces the model estimates $\mu(\bm{p})$. The total number of parameters in vector  $\bm{p}$ is denoted as $n_p=n_1+\dots +n_s$. Then our objective is to minimize 
$$q(\bm{p})=\sum_{k=1}^{d} \max \left\{0,(y_k-\mu_k(\bm{p}))w_k(y_k+u_k-\mu_k(\bm{p}))\right\}.$$

The procedure which leads to an estimates of the coefficients $\bm{p}$ is the following.
\begin{enumerate}
	\item Given an initial guess of the model parameter vector $\bm{p}$, denoted by $\bm{p}^g$, the model equations are numerically solved and model estimates $\bm{\mu}$ are obtained. 
	\item By repeatedly solving the model with slight changes in parameters, we obtain an estimate of the $d\times n_p$ matrix $\bm{J}$ where $J_{ij}=\partial{\mu_i}{\partial p_j}$. We use the approximation
	$$J_{ij}\sim \dfrac{\bm{\mu}_i(\bm{p}^g+\delta_j\bm{e}_j)-\bm{\mu}_i(\bm{p}^g-\delta_j\bm{e}_j)}{2\delta_j},$$
	where $\delta_j$ is a small number and $\bm{e}_j$ are vectors of the canonical basis.
	\item The quantity $\bm{\mu}$ and $\bm{J}$ are used to construct a quadratic model of the fitting objective as a functional of $\bm{p}$
	$$q(\bm{p})\sim \sum_{k=1}^d \max\left\{0, \left(\hat y_k - \sum_{h=1}^{n_p} j_{kh} \bm{p}_h \right) w_{k} \left(\hat y_k +u_k - \sum_{h=1}^{n_p} j_{kh} \bm{p}_h \right) \right\} $$
% 	      $$q(\bm{p})\sim (\bm{\hat{y}}-\bm{J}\bm{p})^T\bm{W}(\bm{\hat{y}}-\bm{J}\bm{p}),$$
% 	      where $\bm{\hat{y}}=\bm{y}-\bm{\mu}+\bm{J}\bm{p}$, and $\bm{W}$ is a diagonal matrix with $W_{kk}=w_k$.
where $\hat y_k=y_k-\mu_k(\bm{p}^g_h)+\sum_{h=1}^{n_p} j_{kh} \bm{p}^g_h$.
	\item We find the new direction to modify $\bm{p}^g$ minimizing, with respect to $\bm{p}$, the quadratic model of the real fitting objective $q(\bm{p})$ together with the constraints. A new value of the parameter vector $\bm{p}^g$ is found.
	\item With the new value $\bm{p}^g$, we iterate steps 1--4 to convergence, obtaining the estimate $\bar{\bm{p}}$.
\end{enumerate}
It is worthwhile to note that this procedure follows the method proposed in \cite{wood2001},  however the quadratic functional to be minimized has been suitably modified to take into account the measurement error in the observations. %As the authors suggest, the method can be improved choosing a different quadratic model of the fitting objective or taking into account an additional term in the objective which is a sum of the ``wiggliness'' measures for the model unknown functions. However, the version proposed in this paper, despite its simplicity, gives satisfactory results. Note also that the particular choice of the model equations (e.g.~partial differential equations, ordinary differential equations) used at the first two steps of the estimation method is not constrained by the method itself.
%%%%%%%%%%%%%%%%%%%%%%%%%%%%%%%%%%%%%%%%%%%%%%%%%%%%%%%%%%%%%%%%%%%%%%%%%%%%%%%%%%%%%%%%%%%%%%%%%%%%%%%%%%%%%%%%%%%%%%%%%%%%%%%%%%%%%%%%%

% \subsection{Mortality estimate on simulated data}
% To check the goodness of the estimation method, we generate a set of data on population abundances using a known form for the mortality rate functions of the four stages:
% $$
% m^i(T)=....
% $$
% Abundances of the different stages are generated starting from the initial condition of ..... and the temperatures recorded at Colognola ai Colli for the year .....

\subsection{Confidence bands}
In addition to the uncertainty in the collected data, there is a variability due to the estimation procedure, that reflects into a variability of the mortality rate functions. 
To account for the effects of the  variability in the mortality rate function, we consider the confidence bands for both the mortality rate functions and the population dynamics. To this end we observe that, denoting by $\bm{\hat p}=\textnormal{argmin}_{\bm{p}} q(\bm{p})$ the estimator, the parameter covariance matrix is given by~\cite{marsili-libelli2003}
\begin{equation}
\label{cov}
C_J(\bm{\hat p})=\frac{q(\bm{\hat p)}}{d-n_p} \left( \bm{J}^T \bm{W}^{-1} \bm{J} \right)^{-1}.
\end{equation}
For large samples, $\bm{\hat p}$ has approximately a multivariate normal distribution with mean $\bm{\bar p}$ and covariance matrix $C_J(\bm{\bar p})$~\cite{wood2001}. The multivariate normal distribution does not ensure to obtain non-negative mortality rate functions. To satisfy this requirement, we introduce a more restrictive constraint drawing the parameter from a truncated multivariate normal distribution.
We draw a certain number of values of the parameter vector, corresponding to different mortality functions which are used to obtain confidence bands for mortality. Then, we run the model for all the mortalities to determine the confidence bands for the population dynamics.

\section{The case of {\em L. botrana}}
In order to estimate the mortality rate functions of the grape berry moth, we need abundance data. We consider two cases. In Subsection \ref{datigenerati} we generate from model \eqref{kolm1}--\eqref{kolm4} a dataset of abundances for the four stages of the grape berry moth to check the convergence of the estimation method. 
% Then, in Subsection \ref{datireali} the method is applied to  field data collected in a Garganega vineyard located in Colognola ai Colli for the years 2008, 2009 and 2011. 
Then, we consider the same dataset used in \cite{gilioli2016} concerning the dynamics of the grape berry moth 
in a vineyard of Garganega located in Colognola ai Colli (Verona, Veneto Region) in the North-East of Italy, during the period 2008--2012. 
The experimental field was not treated with pesticides. 
To estimate the mortality rates functions we used the field data collected at Colognola ai Colli during the three years 2008, 2009 and 2011 (Subsection \ref{datireali}); data collected during other two years (2010 and 2012) were used to validate the model (Subsection \ref{validation}), keeping all the other parameters of development and fecundity fixed. 
In both cases, development and fecundity rate functions are those defined in Section \ref{lobesiabotrana}, and the values of the diffusion coefficients are set $\sigma^i=0.0001,\; i=1,2,3,4$.
Hourly temperature data, collected by a meteorological station close to the vineyard, are used as a driver environmental variable for the model simulation.
The times at which the phenological stage of the plant $P(t)$ reaches the BBCHs  reported in Table \ref{TabPlant} vary over the years; they are reported in Table \ref{time_bbch} for the vineyard of Garganega located in Colognola ai Colli for the years 2008--2012.

\begin{table}
\centering
\begin{tabular}{cccccc}
\toprule
$P(t)$ & 2008 & 2009 & 2010 & 2011 & 2012\\
\midrule
BBCH 53 & 122 & 125 & 126 & 108 & 122 \\
BBCH 71 & 162 & 150 & 158 & 151 & 162 \\
BBCH 81 & 218 & 201 & 209 & 192 & 218  \\
BBCH 89 & 260 & 265 & 274 & 255 & 260 \\
\bottomrule
\end{tabular}
\caption{Days needed from $1^{st}$ of January to reach the BBCH  reported in the first column. The data refers to the vineyard of Garganega located in Colognola ai Colli during the years 2008--2012.}
\label{time_bbch}
\end{table}
Observed abundances are subject to a measurement error depending on the stage. For instance, it is very difficult to detect eggs and pupae. In fact, to count first-generation eggs it is necessary to pick up small grape bunches and to analyse them in laboratory using a stereo microscope. For the other generations a magnifying glass is used to observe bunches in the field and this operation requires particular attention. From the third generation, pupae are under the rhytidome, while second-generation pupae can be hidden either in the grape berries or under the rhytidome, making the detection of this stage particularly difficult. Then, for eggs and pupae we consider a measurement error up to $50\%$ of the collected abundance. Larvae are more visible, then we suppose to have a measurement error up to $10\%$ for this stage. Adults are caught with pheromone traps and we assume they are correctly measured. 

The mortality rate functions are expressed as linear combination of cubic B-splines. The basis is built on the nodes $[0,20,40]$, being $[0,40]$  a suitable interval of temperature, and hence it consists in five polynomials $\xi_{j}(T(t)),\; j=1,\dots,5$ defined on this interval. The mortality rates function are 
$$m^i(T(t))=\sum_{j=1}^{5}p_{ij}\xi_{j}(T(t)), \quad i=1,\dots,4.$$
The constraint minimization of $q(\bm{p})$ (fourth point of the estimation method explained in Section \ref{method}) has been carried out using the \texttt{fmincon} MATLAB routine. In this setting it is possible to include the constraints on the  mortality rates (stated in Subsection \ref{submort}) as linear inequalities involving the parameters $p_{ij}$. %In particular, we choose to test the mortality rates non-negativity/positivity on several meshpoints of the temperature interval $[0,40]$.

\subsection{Generated data}
\label{datigenerati}

\begin{figure}
\includegraphics[width=\columnwidth]{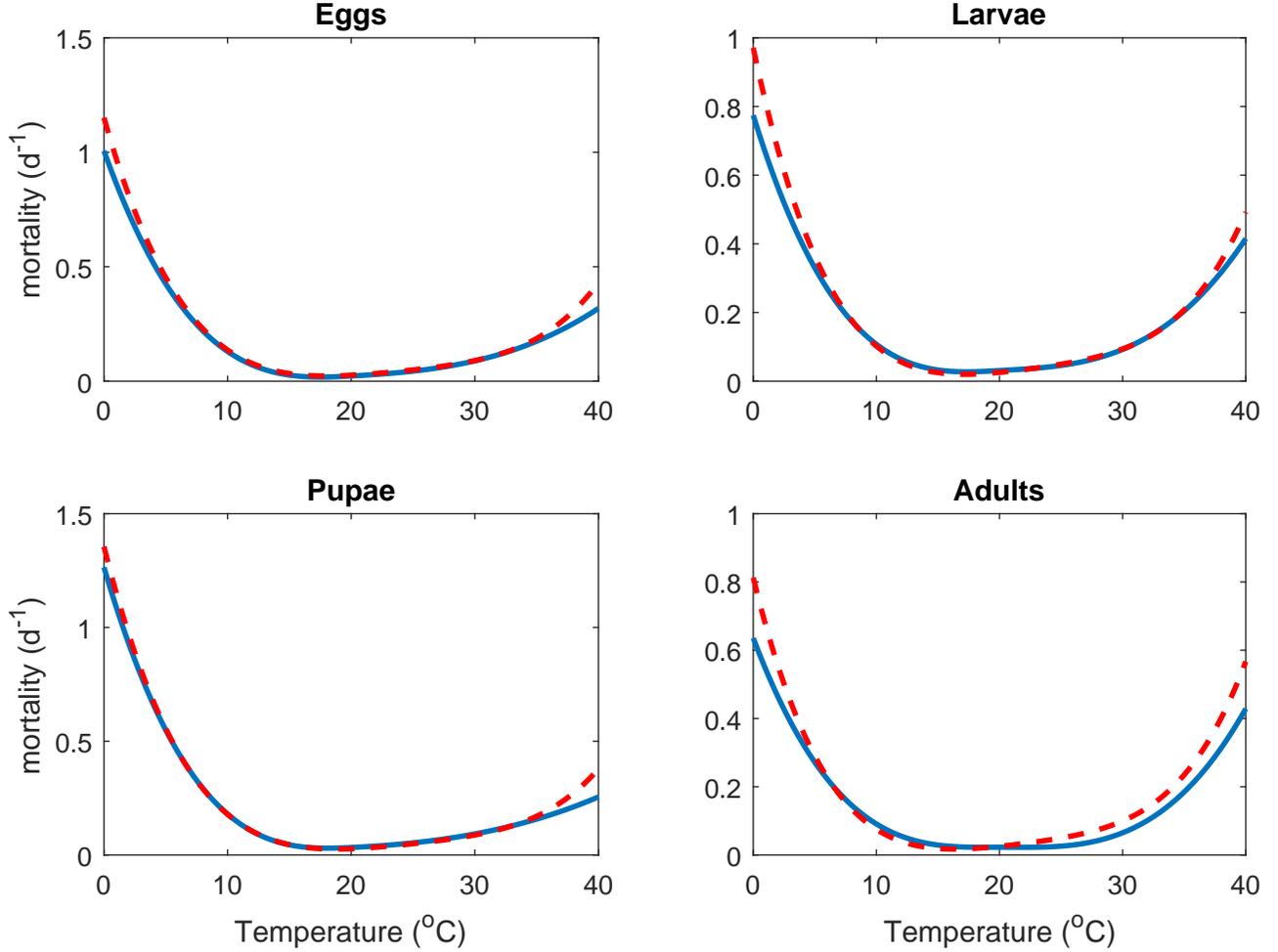}
\caption{Mortality functions for all the stages of the grape berry moth. Red dashed line: bath tube mortality rate functions used to generate data; continuous blue line: estimated mortality.}
\label{mortalita_datigenerati}       
\end{figure}

To check the convergence of the estimation method described in previous subsection, we generate data for three years using the available hourly temperatures , from the field case, for the years 2008, 2009, 2011. We run the model~\eqref{kolm1}--\eqref{kolm4}, using the development and fecundity rate functions defined in Section \ref{lobesiabotrana} with initial condition of 100 adults with physiological age 0 at May $1^{st}$.
The mortality rate functions used to generate data are the so called bath tube shape functions, represented in Figure \ref{mortalita_datigenerati} with a red dashed line.
We pick a value of abundance each 10 days, for all the stages, starting from May $1^{st}$, up to the end of the year.
Then, for each observation, we draw a value $u$ from a uniform distribution over the interval $[0,0.5]$ for eggs and pupae and over the interval $[0,0.1]$ for larvae and multiply the abundances by $1-u$ obtaining the underestimated observations  that will be used in the estimation procedure.

\begin{figure}
\includegraphics[width=\columnwidth]{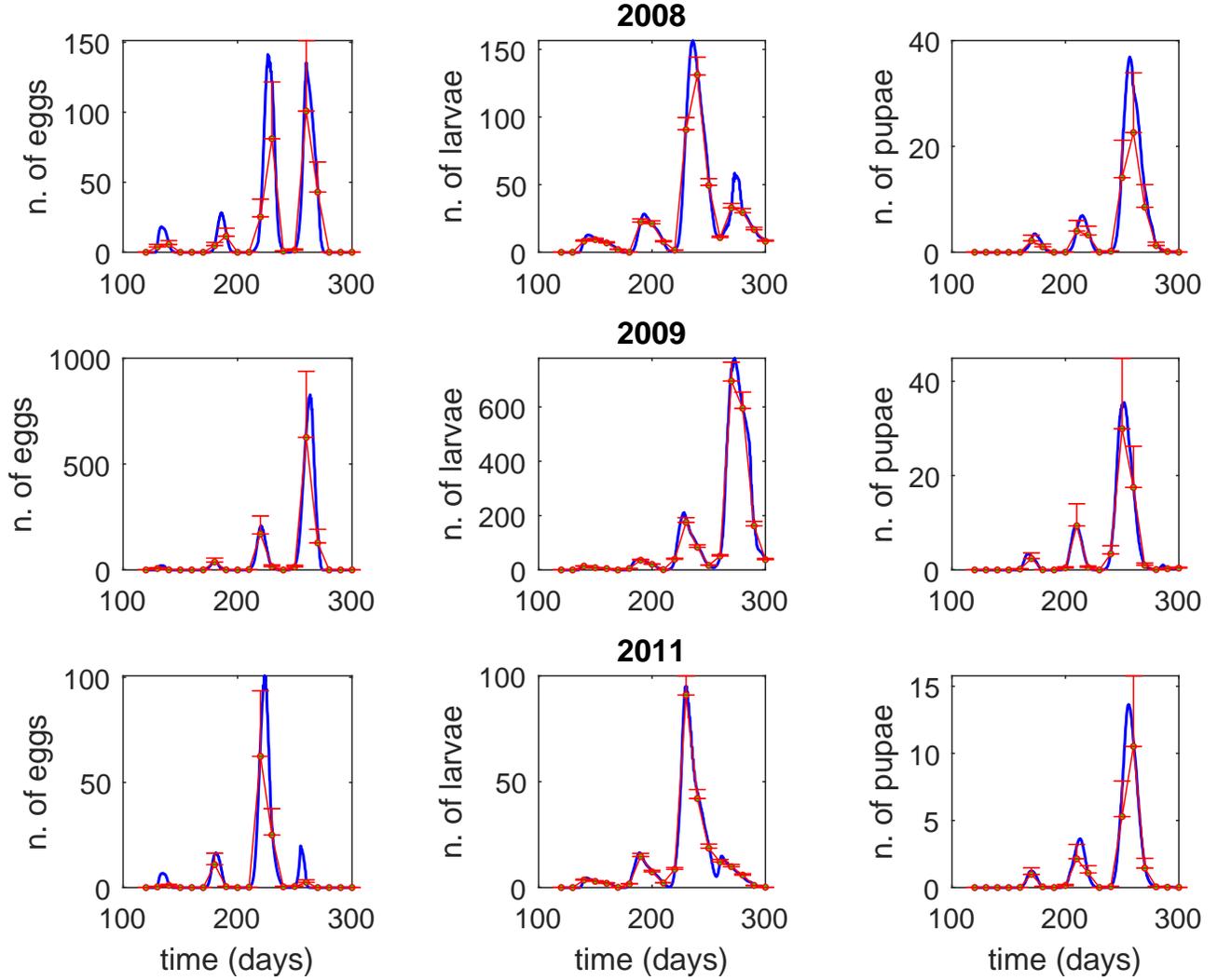}
\caption{Estimation starting from generated data. Red points: data generated from model \eqref{kolm1}--\eqref{kolm4} using hourly temperatures recorded in a Garganega vineyard in Colognola ai Colli for the years 2008, 2009, and 2011. Vertical red intervals represent the range of variability of observed data ($50\%$ of the eggs and pupae abundance, $10\%$ of larval abundance. Blue continuous line: simulated population dynamics obtained with the estimated mortality functions (blue continuous line in Figure \ref{mortalita_datigenerati}).}
\label{dinamiche_datigenerati}       
\end{figure}

To estimate the mortality rate functions, we consider equal weights for all the data. 
We have to minimize the sum of the weighted squared differences $q(\bm{p})$ between simulated dynamics and observations for all the three years 2008, 2009, 2011. More precisely, $d=d_{2008}+d_{2009}+d_{2011}$, where $d_{Y}$ is the number of observation in year $Y$.
Applying the method described in Subsection \ref{method} we obtain the estimated mortality reported in Figure \ref{mortalita_datigenerati} with a blue continuous line.  It can be seen that the estimated mortalities well approximate the bath tube shape functions used to generate data, in particular in the central part of the temperature interval. For low and high values of temperatures some differences between the two curves can be observed. However, looking at Figure \ref{dinamiche_datigenerati}, it is evident that these differences do not produce unpleasant consequences in the population dynamics because low and high temperatures are very unlikely in temperate regions (as Colognola ai Colli). The simulated dynamics, reported in Figure \ref{dinamiche_datigenerati} for the immature stages, fit very well the data, meaning that the simulated trajectories cross the observations interval of variability. This results shows that the algorithm can provide a good approximation of the mortality functions.
We remark that we do not report the confidence bands in this case because they are very small for both mortality and dynamics.

% The simulated dynamics reported in Figure \ref{dinamiche_datigenerati} together with the $95\%$ confidence bands fit perfectly the data. Confidence bands are so small that they are not evident.  This is a further confirmation of the convergence of the estimation method.

\subsection{Field data}
\label{datireali}

In the case of field data, we must have information on the population densities at the beginning of the season to drive the simulation during the entire growing season. In our study, the number of adults caught by a trap and recorded every week since the first larvae of the first generation is observed, were used as the initial condition of the model.
Again, we have to minimize the sum of the weighted squared differences $q(\bm{p})$ between simulated dynamics and observations for all the three years 2008, 2009, 2011, with $d=d_{2008}+d_{2009}+d_{2011}$. As initial guess we used a linear combination of cubic B-splines that approximate the mortality rates already used in~\cite{gilioli2016}.

In the case of field data we do not consider equal weight for all the stages, but we assign a 
%Secondly, we consider a weight matrix $\bm{W}$ giving 
higher weight to the larval stage. This choice relies on the harmfulness of the larvae that in second generation cause severe damage to the grapes. Then, it is important to obtain a good forecast of the larval abundance, in particular to the end of pest control. Moreover, data collected on larvae are more reliable than for the other stages. In particular, the weights assigned to larval observations are one hundred times the weights of the observations assigned to the other stages.
To take care of the effect of the variability in the parameters estimates, we consider the confidence bands for both the mortality rate functions and the grape berry moth dynamics. In particular, we know that the probability distribution of the parameter $\bm{\hat p}$ can be approximated by a multivariate normal distribution with mean $\bm{\bar p}$ and covariance matrix \eqref{cov}. We draw 500 values of the vector parameter from this distribution, that correspond to 500 mortalities for each stage expressed as $\max\{0,\sum_{j=1}^{7} p_{i,j}\xi_{i,j}(t)\}$, because they cannot be negative.
The $95\%$ confidence bands of the mortality rate functions are obtained from these 500 mortalities using the MATLAB routine \texttt{prctile}. To see the effects of the mortality variability in the population dynamics, we run the model for the 500 mortalities computed to obtain the mortality confidence bands and then we determine the population dynamics bands.

\begin{figure}
	\includegraphics[width=\columnwidth]{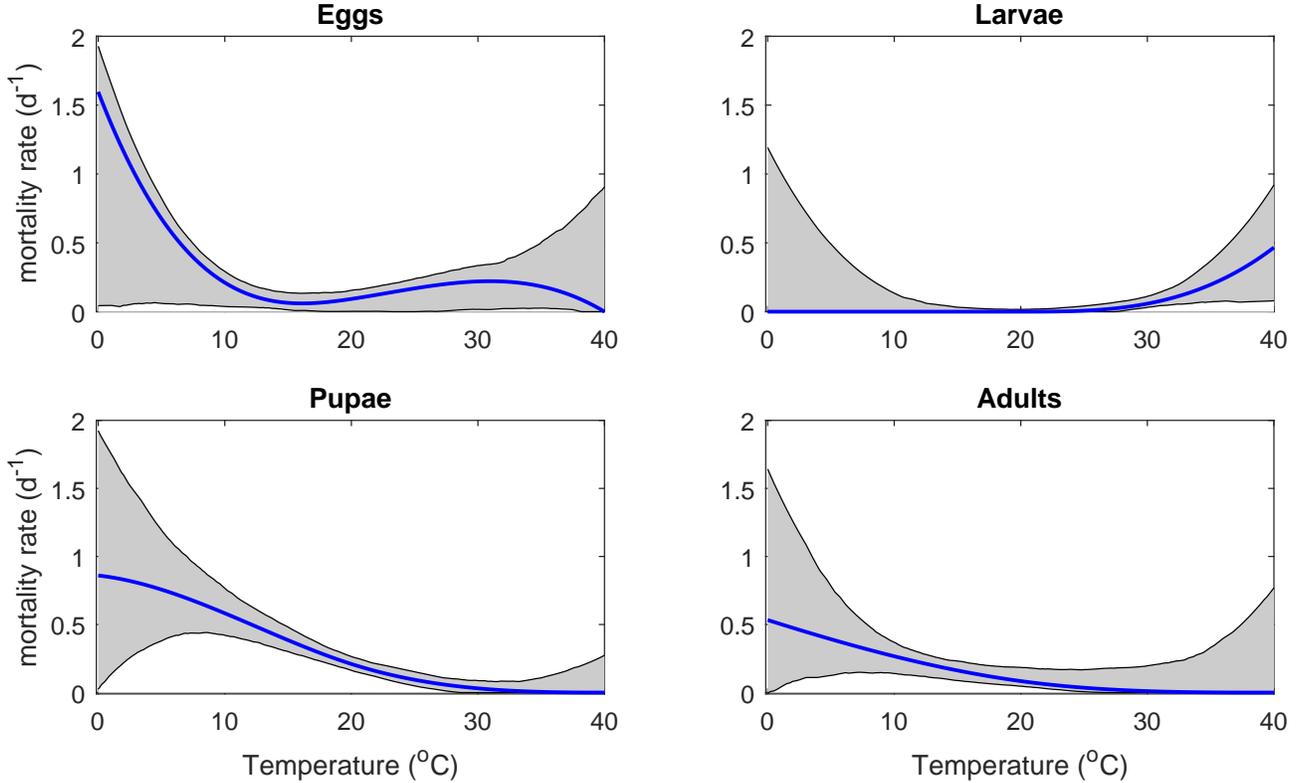}
\caption{Estimated mortality functions from field data and their $95\%$ confidence bands. Blue continuous lines: estimated mortality functions. Grey area: $95\%$ confidence bands for the mortality rate functions of the different stages of the grape berry moth.}
\label{estMor}       
\end{figure}

\begin{figure}
%   \subfigure[]{
\includegraphics[width=\columnwidth]{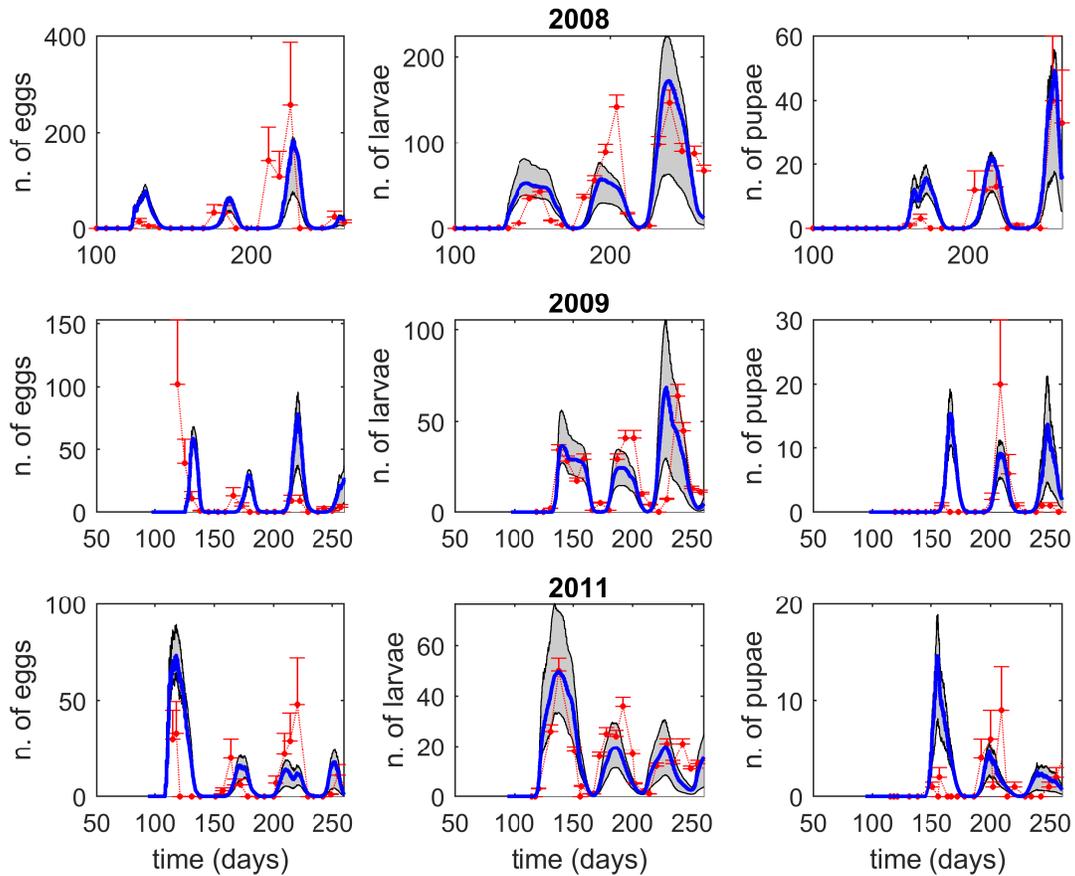}%}
% 	\subfigure[]{\includegraphics[width=\columnwidth]{./immagini/pdynamicsW.eps}}
\caption{Population dynamics in the case of estimation from field data. Red points represent data collected in a Garganega vineyard in Colognola ai Colli for the years 2008, 2009 and 2011, while vertical red intervals indicate the range of variability of observed data ($50\%$ of the eggs and pupae abundance, $10\%$ of larval abundance). The continuous blue lines inside the grey areas show the dynamics obtained using the blue mortalities represented in Figure \ref{estMor}. Grey areas denote the $95\%$ confidence bands for the dynamics of the grape berry moth immature stages.}
\label{dyn080911_pesi}       
\end{figure}

The mortality rate functions obtained in the case of field data and the relative $95\%$ confidence bands are represented in Figure \ref{estMor}. Confidence bands are tight in the central part of the temperature interval, that is when temperatures are favourable to the pest growth. Conversely, they are large at the extremes, that is for unlikely temperatures when no data or few data are available.
The estimated mortalities (blue lines in Figure \ref{estMor}) and the upper bounds of the confidence bands show, for almost all grape berry moth stages, an increasing behaviour for both increasing temperatures greater than an upper threshold and decreasing temperatures lesser than a lower threshold. This behaviour for ``extreme'' temperatures is in agreement with the mortality proposed in~\cite{gilioli2016} where second order degree polynomials were used to describe mortality for low and high temperatures. In the central part of the temperature interval (approximately between $10^oC$ and $30^oC$) mortality has lower values.
As in the previous case of generated data, the variability in the mortality for high and low temperatures does not negatively affect the dynamics of the pest because the mortality values corresponding to these temperatures occur only few times. The pest dynamics for the immature stages are reported in Figure~\ref{dyn080911_pesi}.

% As in the previous case, the mortality function shows, for some stages, an increasing behaviour for increasing temperatures greater than an upper threshold and/or for decreasing temperatures lesser than a lower threshold, while in the central part of the temperature interval (approximately between $10^oC$ and $30^oC$) mortality has lower values, except for pupae. The mortality of pupae is unexpectedly high for likely temperatures; this probably happens to compensate the decrease of the other mortalities in the same temperature interval with respect to the case of equal weights.
% The behaviour for ``extreme'' temperatures is in agreement with the mortality proposed in \cite{gilioli2016} where second order degree polynomials were used to describe mortality for low and high temperatures.

% When considering a higher weight on larvae, the estimated larval mortality is lower than for the case of equal weights, while pupal mortality increases; adult mortality decreases in the central part of the temperature interval and increases for low and high temperatures; also egg mortality increases for low temperatures.
As expected, population dynamics obtained considering a mortality estimated with a higher weight for larvae present a better fit for the larval stage than for the other stages. 
However, the fit for the other immature stages is satisfactory taking into account the large measurement error. 
Looking at Figure \ref{dyn080911_pesi} we observe that for all the three considered years the confidence bands for larval stage include the majority of the observed abundance intervals of variability  or cross these intervals. For eggs and pupae, only in some cases the observed values fall in the $95\%$ confidence bands, but the estimated dynamics allow to obtain a good estimation for larvae. The confidence bands are particularly thick at the peaks of the different insect generations, namely there is a big uncertainty in the maximum abundance.

\subsection{Model validation}
\label{validation}
\begin{figure}
%   \subfigure[]{
\includegraphics[width=\columnwidth,trim={1.5cm 4.8cm 1.5cm 0},clip]{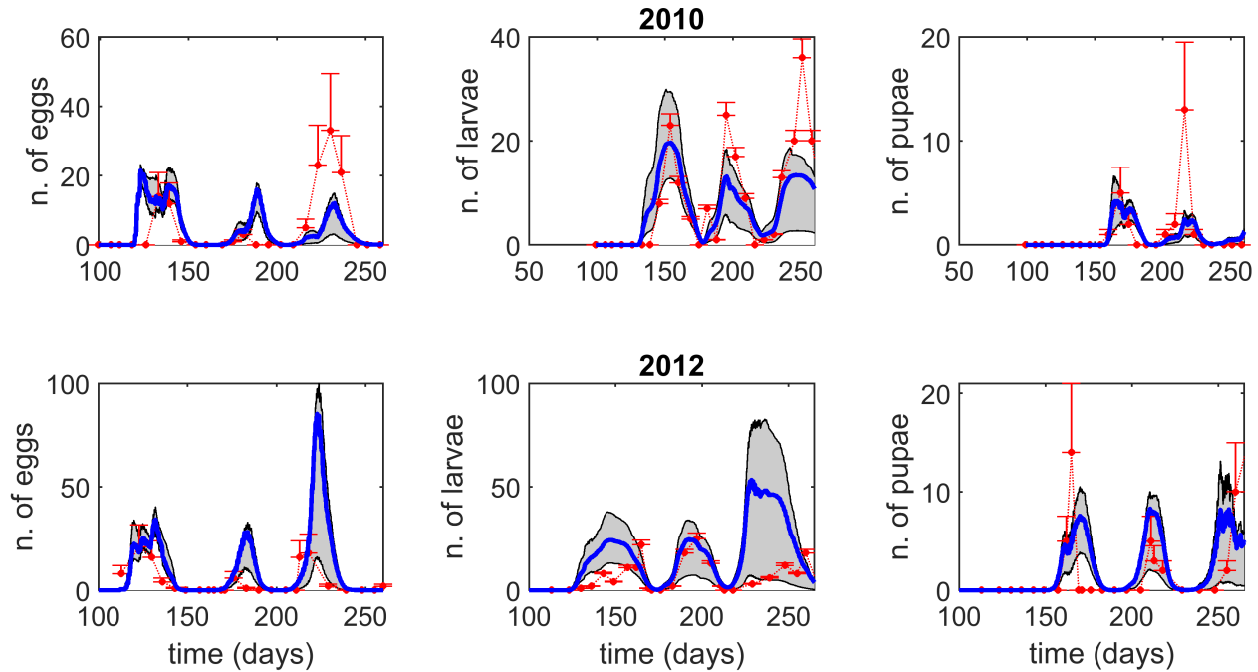}%}
% 	\subfigure[]{
% \includegraphics[width=\columnwidth]{./immagini/dinamiche_garganega_troncamentomortalita2019_conpesi_validazione.eps}%}
% 	\subfigure[2011]{\includegraphics[width=0.6\textwidth]{./immagini/pdynamics2011.eps}}
% 	\subfigure[2012]{\includegraphics[width=0.6\textwidth]{./immagini/pdynamics2012.eps}}
\caption{Validation of the estimated mortality functions. Red points represent data collected in a Garganega vineyard in Colognola ai Colli for the years 2010, and 2012, while vertical red intervals represent the range of variability of observed data ($50\%$ of the eggs and pupae abundance, $10\%$ of larval abundance). Continuous blue lines are the dynamics obtained using the blue mortalities represented in Figure \ref{estMor}. Grey area: $95\%$ confidence bands for the dynamics of the grape berry moth immature stages.}
\label{annisuccessiviNW}       % Give a unique label
\end{figure}

The data for the years 2010 and 2012 recorded in Colognola ai Colli were used to validate the model, keeping all the other parameters of development and fecundity fixed as in the model calibration. The simulated dynamics of the immature stages, obtained using the estimated mortality in Figure \ref{estMor}(a) for the years 2010 and 2012, are represented in Figure \ref{annisuccessiviNW}.
%We report only the case of equal weight for the larval stage; the dynamics in the case of different weights are similar. 
A good representation of the phenology of the species, for the two years considered, is obtained, meaning that the applied method provides a good tool for the mortality estimate.
As in the previous section, many of the observed values or their variability intervals fall into the $95\%$ confidence bands. This means that the estimated mortality rate functions provide a good forecast of the dynamics, in particular for the year 2012.

% \begin{figure}
%   \subfigure[2009]{\includegraphics[width=0.6\textwidth]{./immagini/pdynamics2009W.eps}}
% 	\subfigure[2010]{\includegraphics[width=0.6\textwidth]{./immagini/pdynamics2010W.eps}}
% 	\subfigure[2011]{\includegraphics[width=0.6\textwidth]{./immagini/pdynamics2011W.eps}}
% 	\subfigure[2012]{\includegraphics[width=0.6\textwidth]{./immagini/pdynamics2012W.eps}}
% \caption{Population dynamics obtained with the estimated mortality function showed in Figure \ref{emw} (blue) and compared to field data (red) for years from 2009 to 2012.}
% \label{annisuccessiviW}       % Give a unique label
% \end{figure}

\section{Discussion and concluding remarks}

A realistic simulation of the population dynamics relies on a good knowledge of the biodemographic functions  describing the biology of the species. Frequently, literature data on the mortality rate function are not available and the mortality cannot be easily estimated as the development and the fecundity rate functions. Other methods, based on the availability of population dynamics datasets, have been developed \cite{ellner2002,gilioli2016,lanzarone2017,wood2001}.

Here we consider the case of the grape berry moth, for which we dispose of 5 years of data on population dynamics in the same location: three years are used to estimate the mortality rate functions and two years to validate the  model. 
The observations are affected by a measurement error depending on the stage. To estimate the mortality rate functions of the different stages we propose a method based on the minimization of the sum of weighted squared differences between simulated dynamics and observations. It is based on the method proposed by Wood \cite{wood2001}, but the functional to be minimized is suitably replaced by a new functional able to take into account the variability in the observations.

% In particular, it is very hard to detect eggs and pupae, then we suppose the abundances in these two stages are subject to a $50\%$ errors. Larvae are more evident, then only a $10\%$ error is attributed to this stage.
%To estimate the mortality rate functions of the different stages we propose a method based on the minimization of the sum of weighted squared differences between simulated dynamics and observations. It is founded on the method proposed by Wood \cite{wood2001}, but the functional to be minimized is suitably replaced by a new functional able to take into account the variability in the observations.}

%  to the first group of data. This method returns the estimation of the mortality rate functions corresponding to the best fit of the dynamics for all the three years, meant as the smaller sum of weighted squared differences between simulated dynamics and observations.

The estimation method considered in the present paper has some advantages with respect to the method proposed in \cite{gilioli2016} and \cite{lanzarone2017} for the grape berry moth mortality estimation. In \cite{gilioli2016} and \cite{lanzarone2017} the mortality was represented as sum of two terms: an intrinsic mortality due to abiotic factors and an extrinsic mortality due to biotic factors. The intrinsic mortality was estimated using literature data, while for the extrinsic mortality, estimation methods based on population dynamics observations were  proposed. Here the mortality is considered as a whole and represented as linear combination of cubic splines. This assures a greater flexibility on the shape of the mortality function.

To check the convergence of the estimation procedure, an application to a synthetic dataset where data are generated from system \eqref{kolm1}--\eqref{kolm4} with known mortality rate functions is considered. The observations are then ``corrected'' by an error. The method applied using equal weights for all the stages allows to obtain a very good estimation of both the mortality rate functions and the population dynamics.

Then, we consider the case of field data. We decided to assign a higher weight to larval observations. This is justified by the fact that the larval stage is the most important in pest control since its second generation mainly produces serious damages to the vineyards. Then, it is important to have a good description of larval dynamics. Moreover, the measurements of larvae are more precise than for the other immature stages. In fact, it is very hard to detect eggs and pupae, then we suppose the abundances in these two stages are subject to a $50\%$ errors, while for larvae only a $10\%$ error is considered.

The estimation procedure is subject to variability. In particular, the parameter estimator is a random variable whose distribution can be approximated by a multivariate normal distribution.
Drawing 500 values of the parameter vector from this distribution we can obtain confidence bands for the mortality. 
%The confidence bands again confirm a high uncertainty in the mortality estimate for unlikely temperatures (low and high temperatures), while a small variability is present for likely temperatures (approximately between $10^oC$ and $30^oC$).
Running the model for the 500 mortalities, we obtain the confidence bands also for the population dynamics. The estimated mortalities confidence bands (Figure \ref{estMor}) are thin in the central part of the  interval, for favourable temperatures, %(approximately between $10^oC$ and $30^oC$) 
and large for extreme values of the temperatures showing an increase for high and low temperature values. This behaviour is in agreement with the assumptions made in \cite{gilioli2016} where the mortality is defined as a second order degree polynomial for small and large temperatures that infrequently occur.

%When we assign a higher weight to all larval observations, we observe an increase of the pupal mortality, with respect to the case of uniform weights, and a decrease of the other stages mortalities for likely temperatures. Since the pupae are not very sensitive to temperature, the pupal mortality is unexpectedly high, probably to compensate the effect of the decrease of the other stages mortalities. To avoid this unrealistic behaviour it could be useful to introduce other constraints on mortality rate functions to be used in the estimation procedure.
%Moreover, we observe a decrease in the adult mortality in the central part of the temperature interval, for likely temperatures.

The population dynamics show a good fit of the phenology of the species. Considering a higher weight for larvae, a better fit of the larval stage is observed. Most of the larval observations or their  variability intervals fall into the $95\%$ confidence bands.  This is important in pest control, mainly based on the control of larvae. In this way we can have a satisfactory larval abundance forecast that allows to better plan treatments with pesticides.

%Furthermore, the estimation procedure is subject to variability. In particular, the parameter estimator is a random variable whose distribution can be approximated by a multivariate normal distribution.
%Drawing 500 values of the vector parameter from this distribution we can obtain confidence bands for the mortality. The confidence bands again confirm a high uncertainty in the mortality estimate for unlikely temperatures (low and high temperatures), while a small variability is present for likely temperatures (approximately between $10^oC$ and $30^oC$).
%Running the model for the 500 mortalities, we obtain the confidence bands also for the population dynamics. \textcolor{red}{The variability in the dynamics is more evident in the period of the peaks allowing to include the observations or their variability intervals in the confidence bands.}
%We also remark that it is important to have a good estimate of  the mortality for likely temperatures, while for unlikely temperatures  mortality estimate does not significantly affect the resulting dynamics.

Mortality estimation procedure has been performed considering three years of data on population dynamics. Then, the estimated mortalities have been used to simulate the dynamics for two further years. The satisfactory representation of the phenology in these years allows us to state that the mortality can be actually considered the same for different years (as the development and the fecundity rate functions). This result supports the assumption made in \cite{lanzarone2017} of mortality rate functions constant over the years, and it allows to estimate the mortality once, considering a fixed number of years of observations and then to use the same mortalities for all the following years.

The estimation method here presented is sufficiently general to be applied to other pests. The weighted squared differences to be minimized in the estimation procedure allow to differentiate
the weights for each single observation. In this way it is possible to stress the observations relative to a more sensible stage, as in our case, or the observations of one or more particularly dangerous generations. Moreover, it is possible to quantify the range of variability of the dynamics giving a measure of the error in the forecast of future dynamics.

In conclusion, the approach adds more realism to the stage-structured population model and it enhances its capability to predict population dynamics, which are key issues in developing strategies for pest management.

\medskip
\textbf{Acknowledgements:}
Support by INdAM-GNFM is gratefully acknowledged by CS. 
The authors would like  to thank Tommaso del Viscio for technical support.
%%%%%%%%%%%%%%%%%%%%%%%%%%%%%%%%%%%%%%%%%%%%%%%%%%%%
\bibliographystyle{plain}     
\bibliography{bibliography}
\end{document}